\documentclass[10pt,letterpaper]{article}
\usepackage[top=0.85in,left=2.75in,footskip=0.75in,marginparwidth=2in]{geometry}
\usepackage[utf8]{inputenc}
\usepackage{cite}
\usepackage{nameref,hyperref}
\usepackage[right]{lineno}
\usepackage{microtype}
\DisableLigatures[f]{encoding = *, family = * }
\raggedright
\usepackage{changepage}
\usepackage[aboveskip=1pt,labelfont=bf,labelsep=period,singlelinecheck=off]{caption}
\makeatletter
\renewcommand{\@biblabel}[1]{\quad#1.}
\makeatother
\usepackage{lastpage,fancyhdr,graphicx}
\usepackage{epstopdf}
\fancyheadoffset[L]{2.25in}
\fancyfootoffset[L]{2.25in}
\usepackage{color}
\definecolor{Gray}{gray}{.25}
\usepackage{graphicx}
\usepackage{sidecap}
\usepackage{wrapfig}
\usepackage[pscoord]{eso-pic}
\usepackage[fulladjust]{marginnote}
\reversemarginpar

\begin{document}
\vspace*{0.35in}

\begin{flushleft}
{\Large
\textbf\newline{Modeling dynamics, cell type-specificity and perturbations in Gene Regulatory Networks}
}
\newline
\\
Junha Shin\textsuperscript{1*},
Spencer Halberg-Spencer\textsuperscript{1,2*},
Yuda Liu\textsuperscript{1,2},
Suvojit Hazra\textsuperscript{1},
Erika Da-Inn Lee\textsuperscript{1,2},
Sushmita Roy\textsuperscript{1,2,3}
\\
\bigskip
\bf{$^1$} Wisconsin Institute for Discovery, University of Wisconsin-Madison, Madison, WI, USA, 53715
\\
\bf{$^2$} Department of Biostatistics and Medical Informatics, University of Wisconsin-Madison, Madison, WI, USA, 53726
\\
\bf{$^3$} email: sroy@biostat.wisc.edu
\\
\bigskip
\bf{*} These authors contributed equally to this article. 
\\
\bf{*} This article is scheduled to appear in the \textit{Annual Review of Genomics and Human Genetics}, Volume~19.
\end{flushleft}

\section*{Abstract}
Gene regulatory networks (GRNs) define the regulatory relationships among molecules such as transcription factors, chromatin remodelers, and target genes. GRNs play a critical role in diverse biological processes, including development, disease manifestation, and evolution. However, fully characterizing these networks across multiple cell types and states remains a significant challenge. Recent advances in single-cell omics have dramatically enhanced our ability to measure biological systems at unprecedented resolution. These technologies have opened new avenues for computational methods to infer GRNs, offering deeper insights into cell type-specific mechanisms, causality, and dynamic regulatory processes. This review summarizes the current state of GRN inference from single cell omic datasets, with a particular focus on dynamics and perturbations, and outlines key open challenges that must be addressed to advance the field.

\section*{Introduction}
Gene regulatory networks (GRNs) define connections between regulatory components, such as transcription factors, sequence elements, and target genes, which collectively govern the context-specific, spatio-temporal expression of genes. Advances in single-cell multi-omics sequencing provides a high-resolution view of cellular heterogeneity and cell-type composition across entire tissues and organ systems, and offers unique opportunities to systematically reconstruct GRNs in various biological contexts, including development and disease. Computational GRN inference is essential for understanding fundamental processes such as how stem cells determine their multipotent fates through differentiation in the developmental cell lineage, how differentiated cells can be reprogrammed to a pluripotent state to create patient-specific therapies, and how the cellular regulatory network is disrupted in developmental disorders or cancers. 

\marginnote{
\textbf{Transcription Factor (TF)}\\
a protein that binds to specific DNA sequences to control the rate of transcription of gene
}

Recent efforts in mapping multiple modalities for single-cell data and performing high-throughput perturbation assays offer new opportunities for inferring GRNs. Concurrently, computational methods have advanced to effectively leverage these diverse data types, significantly enhancing our understanding of how GRNs coordinate normal cellular states and contribute to disease mechanisms. Numerous reviews have comprehensively covered the topic of inferring GRNs from single-cell RNA sequencing (scRNA-seq) data. In this review, we aim to explore less-charted territories, with a particular emphasis on modeling dynamics, incorporating cis and trans regulation, and utilizing perturbation data to construct causal networks. Through this focused approach, our aim is to illuminate the nuanced mechanisms underpinning gene regulation and advance the development of more precise and predictive models.

\section*{Typical workflow of inferring GRNs from single cell omic data}
A gene regulatory network (GRN) is a structured framework of interconnected biological components whose regulatory interactions control gene expression dynamics within a cell. These components include transcription factors (TFs), cis-regulatory elements (CREs) such as promoters and enhancers, chromatin remodelers, and the target genes they influence. TFs bind to CREs associated with specific genes to activate or repress transcription. GRNs are further shaped by long-range chromatin interactions and 3D genome organization, which bring distant regulatory elements into physical proximity with their targets. GRNs are often classified into two types \cite{thompson_fungal_2009,zeitlinger_perspective_2024}. cis-GRNs explicitly model the relationship between CREs and gene expression, capturing local regulatory interactions between CREs and their target genes. In contrast, trans-GRNs focus on coordinated expression changes between regulators such as transcription factors and signaling proteins and their downstream targets across individual cells and conditions. Based on this distinction, computational methods for GRN inference are generally grouped into cis-GRN approaches that define CRE-gene connections, and trans-GRN approaches that infer regulator-target relationships. Inferring GRNs from single-cell data typically requires measurements of gene expression and chromatin accessibility, enabling the reconstruction of both cis- and trans-regulatory relationships at cellular resolution.  The integration of these modalities and strategies is further discussed in the section ``Integrating multi-ome datasets to infer cis- and trans-GRNs.''

\marginnote{
\textbf{cis-Regulatory Elements (CREs)}\\
Non-coding regulatory DNA (100 bp–1 kb) sequences which contain TF motifs, including promoters, enhancers, and chromatin elements\\[0.2cm]
\textbf{Enhancer}\\
a non-coding DNA sequence located far from a gene that enhances transcription via transcription factor-promoter interactions\\[0.2cm]
\textbf{Promoter}\\
a non-coding DNA sequence near a gene start site that recruits transcription factors and coactivator proteins to initiate transcription
}

\begin{figure}[htbp] 
\includegraphics[width=\textwidth]{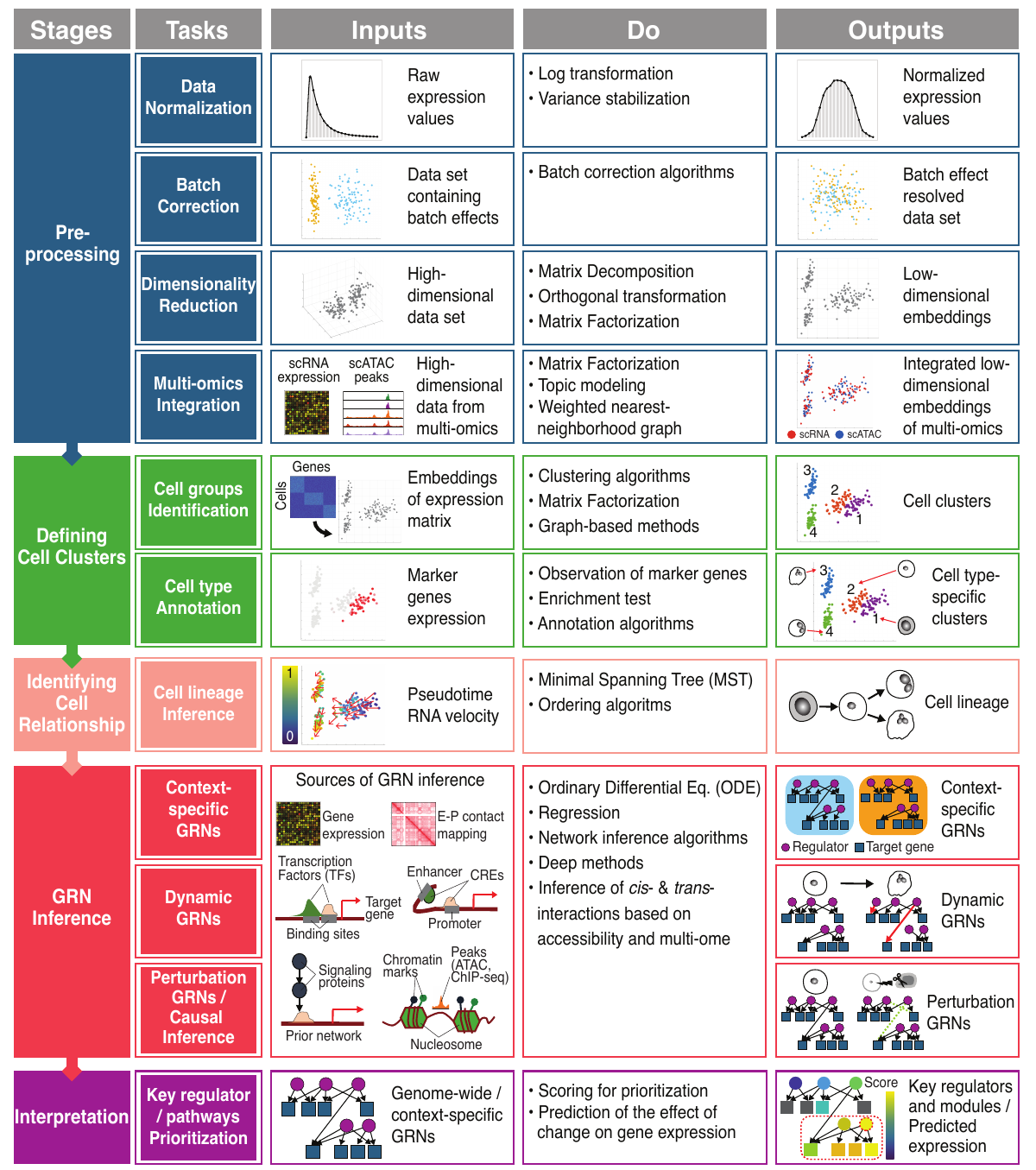}
\caption{A typical workflow of the main stages of inference and analysis of gene regulatory networks (GRNs)  from single cell omics data: Pre-processing, Cell clustering and annotation, Cell-cell relationship  inference, GRN inference and interpretation. Each stage has multiple tasks. Shown are the inputs, computational approaches (Do) and outputs produced by each task. E-P, enhancer-promoter; CRE, cis-regulatory element.}
\label{fig1}
\end{figure}

Prior to the inference of GRNs, several upstream computational tasks must be addressed. These include batch correction, normalization, dimensionality reduction, cell clustering, and reconstruction of cellular relationships through trajectory or pseudotime inference (\textbf{Figure \ref{fig1}}). There are comprehensive reviews that cover these upstream data pro-processing and visualization, and readers interested in these topics are encouraged to consult those sources \cite{flynn_single-cell_2023,liu_single-cell_2016,luecken_current_2019,stegle_computational_2015}. In the following section, we briefly review clustering, cell type annotation, and trajectory inference, as these steps are typically the immediate precursors to GRN reconstruction.

\subsection*{Clustering and data integration}
A key step before inferring GRNs for a dynamic process with discrete cellular transitions is clustering cells to define distinct cell types and states within a single-cell dataset. Cell clustering identifies cellular sub-populations with shared expression profiles through a data-driven approach. Expression or accessibility profiles from single-cell omics measurements usually consist of complex, high-dimensional genetic features. Additionally, expression profiles from multiple modalities cannot be simply combined due to fundamental differences between various biological molecules. Therefore, these profiles are typically simplified and unified through pre-processing steps, including feature selection and/or dimensionality reduction. Feature selection filters out less informative and noisy features to remain only highly variable genes (HVG). Dimensionality reduction maps features into a unified, significantly smaller, lower-dimensional manifold space, highlighting the important patterns and facilitating the integration of datasets across different modalities. These methods include  matrix factorization such as principal component analysis (PCA) and non-negative matrix factorization (NMF), non-linear techniques such as t-distributed stochastic neighbor embedding (t-SNE) and uniform manifold approximation and projection (UMAP), and deep neural network-based methods such as autoencoders \cite{eraslan_single-cell_2019, lopez_deep_2018, rashid_dhaka_2021}. The lower-dimensional representations from dimensionality reduction, i.e. reduced embeddings, enable downstream tasks such as clustering, visualization, and integration of datasets from different sources and modalities.

\marginnote{
\textbf{Non-negative Matrix Factorization (NMF)}\\
a matrix decomposition method that reveals latent structure in non-negative (0 and positive values) data by factorizing into two lower-dimensional non-negative matrices representing row and column features\\[0.2cm]
\textbf{Embedding}\\
a numerical representation of data points in a continuous vector space that captures the properties and the relationships
}

For the identification of cell clusters, inter-cellular distances or similarities are computed based on reduced embeddings, subsequently coupled with unsupervised clustering algorithms such as k-means, hierarchical clustering, or a consensus algorithm of multiple methods \cite{kiselev_sc3_2017}. Alternatively, pairwise similarity between cells could also be modeled as a similarity-based nearest neighborhood graph, subsequently coupled with graph clustering algorithms such as Louvain clustering \cite{macosko_highly_2015, wolf_scanpy_2018}. Numerous data-specific factors and noise impact clustering performance, making it unlikely for any single method to be universally superior \cite{kiselev_challenges_2019, petegrosso_machine_2020}, which makes ongoing evaluation of clustering essential. Among the multiple ways for validating the clustering results, visualization of the cell clusters via t-SNE- or UMAP-based coordinates allows an intuitive comparison between the results from different algorithms or different hyper-parameters of each algorithm.

Single-cell genomic assays are rapidly expanding beyond the transcriptome to other modalities \cite{baysoy_technological_2023}, such as chromatin accessibility with single-cell Assay for Transposase-Accessible Chromatin sequencing (scATAC-seq)\cite{buenrostro_single-cell_2015}, DNA methylation \cite{angermueller_parallel_2016}, 3D genome organization with scHi-C \cite{nagano_single-cell_2013, ramani_massively_2017}, histone modifications with sc-Cleavage Under Targets and Tagmentation (CUT\&Tag) sequencing \cite{bartosovic_single-cell_2021}, and protein abundances \cite{stoeckius_simultaneous_2017}. Although optimally leveraging such multimodal measurements is associated with several unique computational challenges \cite{argelaguet_computational_2021, efremova_computational_2020}, they open up new opportunities to define cell types and states, and their underlying gene regulatory networks, providing a more complete picture of the molecular programs of dynamic biological processes which the transcriptome alone cannot capture \cite{baur_data_2020}. Typically, integration of single-cell transcriptomic data with other modalities involves deriving shared low-dimensional embeddings for each dataset, followed by the application of modality-bridging algorithms to align and merge them. 

These integration strategies can be broadly categorized into gene-centric and feature-distinct approaches. Gene-centric methods, such as UINMF \cite{kriebel_uinmf_2022} and scJoint \cite{lin_scjoint_2022}, rely on fully or partially shared gene features across modalities to align cellular profiles. In contrast, feature-distinct approaches address integration across entirely non-overlapping feature spaces, such as the integration of gene expression and chromatin accessibility regions. These methods employ various strategies, including the use of ``bridge'' cells \cite{lee_integration_2024} that contain measurements from both modalities (multiVI \cite{ashuach_multivi_2023}, Seurat v5 \cite{hao_dictionary_2024}), per-modality dimensionality reduction (scMoMaT \cite{zhang_scmomat_2023}, GLUE \cite{cao_multi-omics_2022}) or computing cell-cell distances independently (MMD-MA \cite{liu_jointly_2019}, SCOT \cite{demetci_scot_2022}), followed by projection into a shared latent space.

\subsection*{Cell type identification}
The identification of cell types within clusters enables researchers to characterize the cellular composition of a population and infer dynamic biological processes. Typically, cell types are annotated by comparing cluster-specific expression profiles to known specific marker genes, often through manual inspection of individual marker gene expression. However, manual annotation is labor-intensive, less reproducible, and fundamentally limited by challenges such as ambiguity of cell types or states and the subjective criteria used to select marker genes. To address these limitations, enrichment analysis of known marker genes among the differentially expressed genes (DEG) of each cluster can be employed, facilitating more systematic and reproducible cell type assignment by revealing overrepresented markers. Moreover, several algorithms have been developed to automate the annotation process \cite{kiselev_challenges_2019,pasquini_automated_2021}. This includes scoring clusters by comparing them to the reference databases from large consortia \cite{kiselev_scmap_2018, zhang_probabilistic_2019}, training a classifier to project cell type identities from known cells to nearest-neighbor unknown cells \cite{pliner_supervised_2019, wang_unifying_2019}, or using sequential binary classifiers informed by structured, ontology-based cell type databases \cite{bernstein_cello_2021}.

For some dynamic systems, such as those with cells in a continuously changing system with high levels of heterogeneity, cell clustering may not be feasible. For these cases, cluster-free approaches based on computational classification \cite{forlin_cluster-free_2024} or cell-/pseudocell-based trajectories could be used, which will be discussed in subsequent sections. Another cluster-free approach is to use a reference dataset or atlas to annotate cells by querying an input set of cells. These methods also assign cell types by identifying the closest matches through the comparison of individual query cell expression profiles to a reference dataset (Azimuth \cite{hao_integrated_2021}), which was collected from large consortia such as HuBMAP \cite{hubmap_consortium_human_2019}, or inferred through deep learning from annotated cell atlases (SCimilarity \cite{heimberg_cell_2025}). Although reference-based approaches offer scalable and consistent annotations, their accuracy can be limited by the diversity and granularity of the cell type reference, potentially leading to underrepresentation of rare cell types, transitional states, or species-specific cell types. Despite these challenges, reference-guided approaches remain a powerful alternative when clustering is impractical.

\marginnote{
\textbf{Cell trajectory}\\
a path that a cell follows as its state changes over time\\[0.2cm]
\textbf{Cell lineage}\\
a tree-based relationship structure of multiple cell types or states that reflects the division of cellular fates\\[0.2cm]
\textbf{Pseudocell}\\
an averaged cell profile created by aggregating gene expression or chromatin accessibility data from a subset of cells\\[0.2cm]
\textbf{Metacell}\\
a subset of highly similar single-cell profiles aggregated into a representative, homogeneous unit\\[0.2cm]
\textbf{Pseudotime}\\
a computationally inferred measurement ordering cells along trajectories by transcriptomic similarity, approximating progression through dynamic processes\\[0.2cm]
\textbf{RNA velocity}\\
an estimated vector of future cellular transcriptional states from splicing ratios, capturing direction and rate of gene expression change
}[-2.5cm]

\subsection*{Cell trajectory and lineage inference}
Single-cell RNA-sequencing measures transcriptomes of thousands of individual cells simultaneously, capturing cells in different stages of a biological process and transitions between these cell states. Such transitions play a key role in both normal processes, such as embryonic development, and disrupted processes, such as cancer. Dynamic transitions can be examined at the level of individual cells, as well as subpopulations of cells, potentially reflecting a cell type, or pseudocells (also called metacells) \cite{ben-kiki_metacell-2_2022, bilous_building_2024, persad_seacells_2022}, which group cells together but are smaller collections than cell clusters. Several metrics and approaches have been developed to estimate such cell states and transitions from scRNA-seq datasets, which  have been comprehensively reviewed and benchmarked in previous studies \cite{deconinck_recent_2021, saelens_comparison_2019, wang_current_2021}. Additionally, Weiler \textit{et al.} \cite{calogero_guide_2023} offers definitions and discussions on lineage, pseudotime, and cell trajectories. 

At the level of individual cells, the most popular metrics are \emph{pseudotime} \cite{trapnell_dynamics_2014} and \emph{RNA velocity} \cite{la_manno_rna_2018}, which can be used to track the progress of a cell along a process. The concept of pseudotime was first introduced by the Monocle approach \cite{trapnell_dynamics_2014}, and estimating pseudotime remains an active area of research, with more than 70 different pseudotime estimation algorithms \cite{saelens_comparison_2019}. Pseudotime can be inferred by sequentially arranging cells based on their similarities in gene expression. Popular pseudotime inference algorithms include Monocle3, Palantir, and Slingshot \cite{qiu_reversed_2017, setty_characterization_2019, street_slingshot_2018}. Monocle3 \cite{qiu_reversed_2017} is a graph-based trajectory inference method, which builds the trajectory that individual cells can take during development by learning principal graph from the reduced dimension space leveraging reversed graph embedding. The graph is used to compute pseudotime as a distance of cell from the root node selected by the user or programmatically \cite{deconinck_recent_2021}. Palantir \cite{setty_characterization_2019} tries to model the landscape of differentiation and characterize continuity in both cell state and fate choice by introducing the concept of ``way points'' and carrying on diffusion on Markov transition matrix. We describe Slingshot below with cell cluster-based methods. 

RNA velocity, introduced by La Manno \textit{et al.} \cite{la_manno_rna_2018} leverages the rate of change in expression estimated from the number of unspliced precursor and spliced mRNAs. RNA velocity can then be used to estimate the pseudotime of the underlying cellular process. Identifying the GRN dynamics associated with velocity or pseudotime could provide key insights into how cells transition along different normal or perturbed trajectories. RNA velocity was estimated by the velocyto method, and since then there have been several methods that estimate velocity \cite{deconinck_recent_2021}. The most common software for inferring RNA velocity is scVelo \cite{bergen_generalizing_2020}, a likelihood-based dynamical model that solves the gene-specific transcriptional dynamics. ScVelo generalizes RNA velocity estimation to transient systems by allowing for different splicing rates for different genes.

Methods for inferring trajectories at the cell subpopulation or cluster level are equally popular and can provide more robust estimates of lineage structure because of the larger number of cells at each state \cite{saelens_comparison_2019}. Such methods typically need clustering as an initial step followed by an approach to link the clusters. The majority of such methods have been graph-based, with Slingshot \cite{street_slingshot_2018} and PAGA \cite{wolf_paga_2019} being two popular approaches. The key difference between the two approaches is that Slingshot represents the cell trajectory with a minimum spanning tree (MST), while PAGA uses graph abstraction of the full cell-to-cell graph to infer more complex graph structure. We note that the Monocle method, which was used to estimate pseudotime, also uses graph-theoretic structures, such as MST and reverse graph embedding. The outputs of the PAGA method can be further processed to estimate pseudotime and the software tool scanpy \cite{wolf_scanpy_2018} provides diffusion pseudotime \cite{haghverdi_diffusion_2016} as a tool to accomplish this task.

\marginnote{
\textbf{Minimum Spanning Tree (MST)}\\
a sub-network that connects all the nodes of the original network without forming cycles and with the minimum possible total edge weight
}

Both methods operating at the single cell or cell population level need user input or an explicit step to estimate the root of the trajectory. Estimating directional information is the key to examining dynamics. As both classes of methods offer distinct advantages, methods have been recently developed to combine RNA velocity, graph-based methods, and pseudotime. One of such popular methods is CellRank \cite{lange_cellrank_2022} and CellRank2 \cite{weiler_cellrank_2024}, which generalizes CellRank. CellRank combines the similarity-based trajectory inference with directional information from RNA velocity to learn directed and probabilistic state-change of the cell trajectories. Unlike other approaches, CellRank automatically infers initial, intermediate, and terminal populations of an scRNA-seq dataset and computes a global map of fate potentials, assigning each cell the probability of reaching each terminal state. Based on inferred potentials, CellRank visualizes gene expression dynamics as cells take on different fates and identifies putative regulators of cell-fate decisions. CellRank takes input as scRNA-seq and pre-computed RNA velocity matrix and generates undirected kNN graph of cells based on gene expression similarity. The graph is used to devise a Markov chain with the transition probabilities of cells informed by both transcriptomic similarity and RNA velocity. Finally, graph abstraction is used to decompose the graph into clusters that represent different cell states, types, or stages of a dynamic process. 

In a subsequent section, we discuss different GRN inference algorithms that incorporate different types of trajectories (e.g., pseudotime, velocity, trees, and graphs) to gain better insight into the fine grained dynamics of a biological process.

\section*{Modeling regulatory dynamics in GRNs}
\begin{figure}[htbp]
\includegraphics[width=\textwidth]{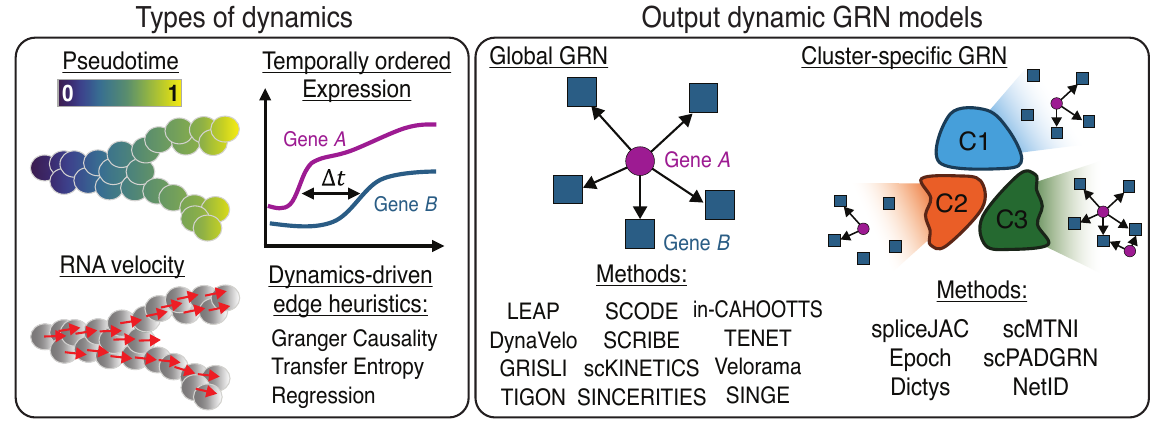}
\caption{\textbf{Inputs and outputs for modeling dynamics in GRNs}. Left box: Types of dynamics among cells inferred using single cell omics data: Pseudotime and RNA velocity, used for input into GRN inference algorithm. Either input can order cells on linear or branching trajectories. Key modeling techniques for GRN dynamics: Granger causality, transfer entropy, statistical regression, which rely on temporal ordering of cells. Right box: Different types inferred dynamics in GRNs. ``Global GRN'' methods infer a single network structure with a single dynamical model capturing expression of target gene as a function of expression of a regulator using expression from a previous cell. ```Cluster-specific GRN'' each cell cluster can have a different network structure. Within each cluster methods may or may not additional dynamics among individual cells within a cluster. Circular nodes: transcriptional regulators; square nodes: target genes.}
\label{fig2}
\end{figure}

The goal of modeling regulatory dynamics in GRNs is to understand how interactions between regulators and genes change over time and influence cellular processes. Edges in dynamic GRNs capture temporal or pseudo-temporal relationships in gene expression levels of regulators and target genes. Here we focus on methods for trans-GRN inference since these are the primary type of methods focusing on dynamics. These methods make different modeling assumptions and have additional inputs such as pseudotime, RNA velocity, cell type specificity constraints, and lineage structure. Methods to obtain temporal inputs can vary between the GRN inference algorithms, but inferred GRNs are presented as a single or multiple cell type-specific network connecting regulators to target genes (\textbf{Figure \ref{fig2}}). 

\subsection*{Incorporating dynamics in GRN based on pseudotime}
The initial approaches that incorporated dynamics into GRN inference from single cell RNA-seq data utilized pseudotime and assumed pairwise dependencies to relate regulators to targets. Among the first methods was LEAP \cite{specht_leap_2017}, which used maximum correlation between gene pairs to infer regulatory interactions by optimizing across different lag times. Although LEAP aimed to resolve temporal relationships in gene expression using pseudotime, it relied solely on pairwise gene co-expression, which is limited in uncovering causal relationships. Another class of methods is based on ordinary differential equations (ODEs) to describe regulatory network  dynamics, as implemented in SCODE \cite{matsumoto_scode_2017}, spliceJAC \cite{bocci_span_2022}, scPADGRN \cite{zheng_scpadgrn_2020}, scKINETICS \cite{burdziak_sckinetics_2023}, Dictys \cite{wang_dictys_2023}, and NetID \cite{wang_scalable_2024}. ODEs describe continuous time dynamics and the underlying physical phenomena, making them suitable for inferring GRNs from scRNA-seq assays during dynamic processes such as differentiation. ODEs can also capture more complex interaction patterns by allowing multiple regulators to influence the expression of the target gene. Practically, ODE-based methods need as input a pseudo-time stamp  and then solve a regression problem using a series of linear equations to represent these changes based on the current expression state of the cells.  

Although ODE-based models are effective for modeling GRN dynamics, most have used linear relationships between regulators and target genes. Recently, linear ODE models have been extended to non-linear models, based on Recurrent Neural Networks (RNNs, Jackson \textit{et al.} \cite{jackson_simultaneous_2023}) and neural ODEs (in-CAHOOTTS \cite{beheler-amass_dynamic_2025}, DynaVelo \cite{karbalayghareh_deep_2025}). In particular, Jackson \textit{et al.} used  RNNs and a prior TF-target network together with very high-resolution yeast time-series scRNA-seq data to estimate transcription synthesis and degradation rate parameters as well as a refined GRN structure. The RNN model assumes that the dynamics evolve in discrete time steps and requires observed expression at each time point. Jackson et al applied this model to a finely resolved yeast time course at 10 minute intervals for 1 hour. The final TF-target network from the RNN-trained model was obtained using TF ablation per gene and using the change in prediction error of a target gene. The rate parameters were also used to estimate the RNA velocity, providing an alternative approach to  spliced/unspliced ratio-based estimates of velocity. Compared to a linear model, the RNN-based TF-target relationships were more accurate based on known regulatory relationships and also predicting expression under regulator perturbation. IN-CAHOOTS uses scRNA-seq alone, while DynaVelo uses both scRNA-seq and scATAC-seq to predict velocity and infer GRNs. The neural ODE model enables a continuous time modeling of the dynamics and does not require measured expression at each time point and can generalize to future time points. Both approaches use a model ablation or gradient-based approach to infer interpretable TF-target relationships. 

\marginnote{
\textbf{Transfer entropy (TE)}\\
a statistical method from information theory that detects causal relationships by measuring directed information transfer between time series or processes using entropy-based computations\\[0.2cm]
\textbf{Granger causality test}\\
a statistical hypothesis test that detects causal relationships by assessing whether one time series improves forecasting of another, using linear regression modeling of stochastic processes\\
}

Information-theoretic methods offer an alternate approach to model non-linear TF-target relationships. One such measure is transfer entropy, a type of conditional mutual information used in methods such as SCRIBE
\cite{qiu_inferring_2020} and TENET \cite{kim_tenet_2021}.  SCRIBE additionally studied  various types of dynamic sources for cell ordering, including pseudotime, RNA velocity, and live imaging. Interestingly, they observed poor performance with live imaging or pseudo-temporally ordered single-cell datasets, which was due to the loss of temporal coupling between measurements of interacting genes, where fluctuations in the levels of a regulator propagate to measurements of its targets. This performance was dramatically improved by using RNA velocity, which suggests that RNA velocity can provide higher quality measurements for temporal ordering of cells in GRN inference.

Several methods additionally make causal assertions between a regulator and target genes based or Granger causality  \cite{barnett_granger_2009} or conditional mutual information, by leveraging the pseudotemporal ordering of cells. 
Granger causality has been frequently utilized for dynamic GRN inference.   For example, both SINCERITIES \cite{papili_gao_sincerities_2018} and SINGE \cite{deshpande_network_2022} use the notion of Granger causality by using regression-based models. Methods using Granger causality often assume uniform spacing of pseudotime which might be violated due to imbalances in the distribution of cells in the latent space. SINGE attempts to overcome this by using a generalized lasso Granger test, which can improve stability in pseudotime periods with low cell numbers. On the other hand, TENET \cite{kim_tenet_2021} and SCRIBE \cite{qiu_inferring_2020} use conditional mutual information theoretic measures like transfer entropy (TENET) or directed information (SCRIBE), that like Granger causality are based on the idea that a causal regulator of a gene should have more information about the future state of a variable  than the previous state of the variable itself. However, the information theoretic framework allows these models to capture non-linear relationships. 

RNA velocity has been used to inform GRN dynamics using other modeling frameworks, such as ODE approaches as well. Because RNA velocity estimates may be incomplete due to the unavailability of splicing information for some genes \cite{bergen_generalizing_2020}, some methods use this only as a starting point to either update RNA velocity or estimate the cellular manifold using conceptually similar quantities like velocity. One of the earliest methods in this area is GRISLI \cite{aubin-frankowski_gene_2020}. This ODE-based regression framework infers a single static GRN and simultaneously infers a velocity vector field based on input scRNA-seq expression and temporal information such as pseudotime or real-time information. TIGON \cite{sha_reconstructing_2024} infers cell velocity based on the cell manifold and multi-time point expression data using optimal transport (OT) and neural networks. Specifically, TIGON assumes that cells at each collected time point have a different distribution and uses OT to map cells between successive time points and estimate velocity using neural networks. This results in cell-type-specific GRNs with dynamic trajectories, although the GRN is significantly small. Velorama \cite{singh_causal_2024} is a Granger causality-based GRN inference method with cell velocity inference that assumes cells on a manifold can be  ``partial ordered''  as a directed acyclic graph (DAG). It uses a neural network to predict the expression levels in a cell based on various conditions of gene regulations simulated upon the DAG structure. 

\marginnote{
\textbf{Optimal transport (OT)}\\
an optimization approach to map  a distribution into another locations by minimizing the cost of transportation
}
 
Another ODE-based methodology, scKINETICS \cite{burdziak_sckinetics_2023}, defines cell velocity more generally based on changes in expression as estimated by an inferred GRN. The GRN structure is then estimated from the new velocities, invoking an Expectation-Maximization style algorithm to estimate both quantities. This algorithm also uses ATAC accessibility information to constrain the network's structure. The GRNs generated by scKINETICS are cell-type-specific, and cell velocity and per-gene velocity are also produced as results.

Overall, a variety of temporal modeling frameworks are available, with methods estimating the cellular manifold being the most powerful, but also most computationally expensive to estimate. Furthermore, methods which use RNA-velocity as a starting point and update it as part of the learning algorithm (e.g., DynaVelo, GRISLI), may be most powerful for modeling GRN dynamics. 

\subsection*{Cell-type-specific GRNs}
The aforementioned algorithms that use pseudotime typically predict a single GRN, exceptions are scKinetics and Dictys, with edges describing the dynamics arising during a temporal process, that is, they make a ``stationary'' assumption that the structure and parameters of the GRN do not change with time. These GRN models are unable to account for gene expression variation arising from state- or cell-type-dependent regulatory interactions \cite{stumpf_inferring_2021}. To capture the complexity and heterogeneity of cell-type-specific gene regulation, a new class of inference algorithms has been developed. These methods use groups of cells representing a cell type or state defined computationally or experimentally with cell-type-specific markers.  Some methods additionally leverage the cell trajectory \cite{zhang_inference_2023}, which can be  inferred using minimum spanning tree-based algorithms such as Monocle \cite{trapnell_dynamics_2014} and Slingshot \cite{street_slingshot_2018}. Methods vary based on the extent to which they model within and between cell cluster dynamics. 

spliceJAC \cite{bocci_span_2022} takes predefined cell annotations as input cell states and models local mRNA splicing dynamics using an ODE-based regression framework to infer cell state-specific gene regulatory interactions. Similarly, scMTNI \cite{zhang_inference_2023} also takes predefined cell type-specific clusters and lineage information derived from uni (scRNA-seq alone) or multimodal single-cell datasets (scRNA-seq and scATAC-seq) and employs a probabilistic graphical model (PGM) within a multi-task learning framework to model the regulatory interactions unique to each cell type. In Epoch \cite{su_reconstruction_2022}, clusters (referred to as `epochs') can be defined based on clustering algorithms or as uniform-sized segments in pseudotemporally ordered cells. It first constructs the holistic static network based on statistical network modeling, powered by prior knowledge from signaling networks, then derives the dynamic GRNs for epochs by refining the static GRN through filtering implausible edges. Several algorithms also model within cell cluster/group dynamics typically leveraging pseudotemporal ordering of cells. For example, scPADGRN \cite{zheng_scpadgrn_2020} performs pseudotime-based clustering and lineage inference using Monocle, and infers cell state-specific GRN using an ODE-based regression framework, enabling smooth changes in dynamic GRNs across states. Dictys \cite{wang_dictys_2023} is another methodology that incorporates scATAC-seq data like scMTNI, using temporal binning of cells along the trajectory to define cell stages. It first infers a transcription factor (TF) binding network using scATAC-seq data, then refines the network with scRNA-seq data to derive dynamic GRNs through probabilistic modeling for each bin. NetID \cite{wang_scalable_2024} uses `metacells' \cite{bilous_building_2024}, defined as a small group of cells exhibiting a similar expression state on a cellular manifold as defined a pruned knn graph. Metacells are ordered and grouped into lineages based on pseudotime or RNA velocity using methods like CellRank or Palantir. Finally, lineage-specific GRNs and their key regulators are inferred by GENIE3 \cite{huynh-thu_inferring_2010} to obtain a ``base'' GRN, and refined into lineage-specific dynamics using  Granger causality.

\subsection*{Modeling per-cell dynamics and GRNs}
Recently, GRN inference methods have focused on estimating (or using) changes in expression profiles, or RNA velocity by modeling per cell GRNs on the cell manifolds. A cell manifold is a high-dimensional representation of cellular states consisting of genes (gene space), where each cell's position reflects its individual gene expression pattern, capturing the continuous spectrum of gene expression profiles across different cells. 

To address the challenging inference task, these models commonly assume that gene expressions of nearby cells along a trajectory are similar and that similar gene expression patterns are produced by GRNs with similar features. In LocaTE \cite{zhang_learning_2023}, the Markov chain of cellular states models the probabilistic transitions of gene regulatory interactions within the cellular neighborhood, expressed as transfer entropy. CeSpGRN \cite{zhang_cespgrn_2022} employs a kernel function to regularize GRNs from similar cells over a smoothed cell-cell graph using a probabilistic graph model. Notably, since both methods learn their own cell manifolds, they do not require additional temporal information such as pseudotime or trajectory inference, nor do they rely on predefined cell clusters. 
 
The main advantage of modeling per-cell GRNs is the fine-grained resolution it offers, enabling the identification of transient cellular states and uncovering nuanced regulatory changes that might not be apparent at a cell cluster level. However, these methods are inevitably prone to false positive regulatory interactions due to their high sensitivity. In contrast, methods that  infer cluster-/state-specific GRNs tend to be more robust and less sensitive, as the trajectories are more stable. Additional benchmarks are needed to assess the value of estimating per-cell GRN dynamics versus more coarse grained dynamics such as at the cluster or entire population level.

\section*{Integrating multi-ome datasets to infer cis- and trans-GRNs}
\begin{figure}[htbp]
\includegraphics[width=\textwidth]{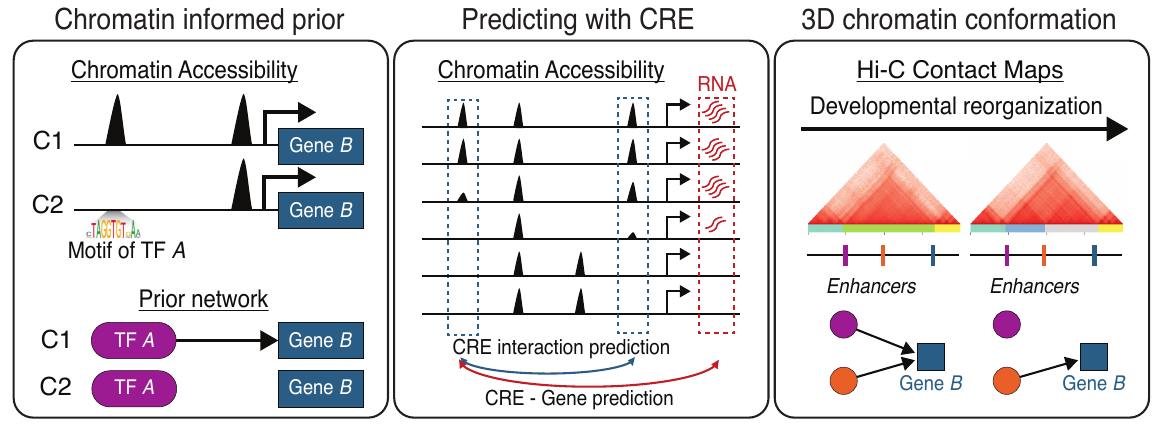}
\caption{\textbf{Integration of multi-omics datasets for cis and trans-GRN inference}. Various methods for integrating chromatin accessibility in gene regulatory network (GRN) inference. Chromatin Informed prior: use of peaks from chromatin accessibility assays to guide structure of trans-GRNs. Prediction with CRE: inference of CRE-gene interactions using correlation or regression models predicting target gene expression from CRE accessibility. 3D chromatin conformation: Use of long-range interactions between CREs and genes to inform GRN structure. Circular nodes: transcriptional regulators; square nodes: target genes.}
\label{fig3}
\end{figure}

The availability of single cell multi-modal datasets \cite{baysoy_technological_2023, chen_high-throughput_2019, kandror_enhancer_2025, ma_chromatin_2020, mimitou_scalable_2021, pedrotti_emerging_2024, vandereyken_methods_2023} has fueled new methodological approaches for GRNs at both cis- and trans-levels. These methods  can be grouped into primarily cis-GRN, primarily trans-GRN, and integrated cis-trans GRN inference methods (\textbf{Figure \ref{fig3}}).

\subsection*{Cis-GRN inference}
A popular approach to infer cis-GRNs from integrated scATAC-seq and scRNA-seq measurements is to use pairwise correlations between a regulatory region and a gene within a particular genomic distance e.g., FigR \cite{kartha_functional_2022}). These methods assume that a regulatory region and gene pair interact independently of other regions or gene pairs. When scATAC-seq and scRNA-seq measurements are not available for the same cell, such methods need to link cells with scRNA-seq measurements to cells with scATAC-seq measurements. Approaches to integrate multi-omic datasets to infer cell clusters, which typically rely on a dimensionality reduction step, such as Canonical Correlation Analysis (CCA), are relevant here. After dimensionality reduction,  a kNN graph-based approach is used to define cell pairs, e.g., scOptMatch in FigR \cite{kartha_functional_2022}. After defining these cell pairs, FigR infers cis-GRNs correlating genomic loci with the neighboring gene's expression profile between computationally matched cells.

More recent cis-GRN inference methods leverage paired scATAC-seq and scRNA-seq  datasets where these modalities are inferred in the same cell e.g., 10x genomics multiome \cite{swanson_simultaneous_2021}, SHARE-seq \cite{ma_chromatin_2020}, DOGMA-seq \cite{mimitou_scalable_2021}). Two methods that leverage this technology are SCENT \cite{sakaue_tissue-specific_2024} and SCARLink \cite{mitra_single-cell_2024}. Both methods assume that the expression of a gene is a function of the activity of  regulatory regions in its neighborhood and cast the cis-GRN inference task as expression prediction of an individual gene with Poisson regression. Both also further assume cis-GRNs are cell-type-specific and infer these regression models per cell type. The main difference between the two models is that SCARLINK considers multiple CREs at a time in an L2-regularized regression model, using all tiles within a 250kb radius of a gene followed by Shapley-based interpretation analysis to identify candidate enhancers per gene. In contrast, SCENT is a pairwise model, considering one gene-CRE pair at a time and uses a non-parametric statistical test applied to each gene and peak pair, with the regression model incorporating additional co-variates including UMIs per cell, mitochondrial reads, and batch. CREs are associated with genes based on a bootstrap confidence score. Compared to existing methods, these approaches predicted expression better and more accurately inferred experimentally supported long-range interactions and exhibited a higher enrichment of causal variants from eQTL studies.

While methods combining scATAC-seq and scRNA-seq can link cell-type-specific CREs to their target genes,  it is still a challenge to accurately predict which distal CREs drive changes in gene expression, especially at longer distances (e.g. beyond 200kb). 3D genome conformation assays, such as Hi-C and HiChIP \cite{dekker_spatial_2023} can provide additional information about candidate regions that interact with gene promoters. Approaches for predicting gene expression, largely based on cell type-specific bulk assays, have been developed that integrate sequence, epigenomic signals, and 3D chromatin architecture data to predict gene expression. Examples of such models include EPInformer \cite{lin_epinformer_2024} and GraphReg \cite{karbalayghareh_chromatin_2022}, which learn a single predictive model of expression. These methods leverage statistically significant loops between pairs of genomic loci,  modeled via a  transformer \cite{lin_epinformer_2024}, or graph neural networks \cite{karbalayghareh_chromatin_2022}. Enhancer-promoter interactions are defined based on ``feature attribution'' \cite{karbalayghareh_chromatin_2022} or ``attention mechanism'' \cite{lin_epinformer_2024}, or a combination of both. As single cell datasets measuring 3D genome organization together with other assays such as DNA methylation and  gene expression emerge \cite{liu_linking_2023, wu_simultaneous_2024, zhou_gage-seq_2024}, a direction of future work is to extend these models for such datasets to improve our understanding of cell type-specific long-range gene regulation.

\subsection*{Trans-GRN inference}
The second class of approaches, which integrate scATAC-seq and scRNA-seq data are for trans-GRN inference. These approaches use the scATAC-seq data to derive an initial TF-target network, defined as a ``prior'', to inform the final scRNA-seq-based GRN and, in some cases, estimate Transcription Factor Activity (TFA) defined as the overall regulatory effect of a TF on its target genes \cite{hecker_computational_2023}. A common strategy for generating these prior networks is to utilize motif instances in the gene promoter, typically defined as a  window $\pm$ transcription start site (TSS). Methods such as scMTNI \cite{zhang_inference_2023}, Symphony \cite{burdziak_nonparametric_2019} and Inferelator \cite{skok_gibbs_high-performance_2022} treat this information as ``soft'' prior to guide which TF-target edges could be added or removed in the final inferred GRN, while others like CellOracle \cite{kamimoto_dissecting_2023}, Pando  \cite{fleck_inferring_2023},  Dictys \cite{wang_dictys_2023} treat the TF-target prior network as the universe of possible interactions and  remove edges not supported by expression. In all these approaches, the mRNA of the TF is used to predict the mRNA level of the target gene using a regularized regression framework. While approaches that use the TF-target network as a soft prior can add TFs and regulators absent from the prior network (provided they explain expression), the other approaches can use only TFs with known motif information contained in the prior. Soft prior frameworks are more flexible, allowing prediction of regulatory relationships that may only be supported by the expression data alone, which could be important if the scATAC-seq data is sparse. 

Many of these methods additionally infer cell-type-specific trans-GRNs, for example, CellOracle \cite{kamimoto_dissecting_2023}, Pando \cite{fleck_inferring_2023}, scMTNI \cite{zhang_inference_2023}, Dictys \cite{wang_dictys_2023}, Inferelator's AMuSR method \cite{skok_gibbs_high-performance_2022}. Some methods infer the GRN for each cell cluster separately (CellOracle, Dictys), while others like scMTNI and AMuSR assume the underlying cell types are related with similar GRNs and use a multi-task learning framework to allow  sharing of regulatory relationships between different clusters. scMTNI further assumes that cell types that are closer on the lineage have similar GRNs and leverages the lineage tree to incorporate this similarity in its multi-task learning framework. Most of these methods assume that the trans-GRNs are influenced by activity around the gene promoters (e.g., $\pm$5000bp around the gene TSS). This assumption limits these methods ability to incorporate regulatory information from longe-range CRE-gene interactions. Furthermore, CREs are incorporated implicitly through the prior, making it difficult to study the direct contribution of trans and cis-factors to a target gene's expression variation. 

\marginnote{
\textbf{Transcription Factor Activity (TFA)}\\
an estimated measure of a transcription factor’s regulatory effect, inferred from expression changes of its known or likely target genes, including post-translational influences
}

In addition to using the prior to constrain the structure of the GRNs, numerous methods use the scATAC-seq data to estimate the ``activity level'' of TFs, or transcription factor activity (TFA) \cite{hecker_computational_2023}. This can be helpful when the TF is post-transcriptionally regulated and its mRNA level may not capture its activity involved in regulating their target genes. TFA is typically estimated independently, upstream of GRN inference and subsequently used together with mRNA levels of TFs for GRN inference. One way to estimate TFA is with ChromVar, which computes a ``deviation score'' defined as a biased-corrected measure of the accessibility of the TF's binding sites in a cluster of interest, relative to a set of background peaks accessible throughout the dataset. The ChromVar score has been used to determine TFs whose accessibility is specific to certain cell states.  

A more common approach for TFA estimation, especially for GRN inference methods is to use a matrix factorization approach, also referred to as ``Network components analysis'' (NCA)\cite{liao_network_2003}. Some methods use  Bayesian probabilistic matrix factorization, such as PMF-GRN\cite{skokgibbs_pmf-grn_2024} and BITFAM \cite{gao_bayesian_2021}. These approaches use gene expression and a prior TF-target connectivity matrix to infer TFA, sequencing depths, and the existence and magnitude of TF-target interactions \cite{skokgibbs_pmf-grn_2024}. As an autoencoder-based approach, SupirFactor \cite{tjarnberg_structure-primed_2024} uses the prior connectivity matrix to constrain its input weights and interprets the first latent activation as the TFA and the output weights as the inferred GRN. An optional hidden layer can be added to aggregate the TFA into an activation that models higher-order TF-TF interactions and infers pathway activity. Several of these approaches output the GRN directly as part of the TFA estimation process, while others perform additional GRN learning treating the TFA as additional predictor variables \cite{siahpirani_uncovering_2025, skok_gibbs_high-performance_2022}. The latter is more general as it can include TF regulators with known as well unknown sequence specificity and can also include additional regulators such as chromatin remodelers and signaling proteins.

\subsection*{Simultaneous cis and trans-GRN inference}
An emerging direction for GRN inference with multi-omic data is simultaneous inference of cis- and trans-GRNs. 
Such models can incorporate distal cis-regulatory interactions, improving sensitivity beyond trans-only GRN inference methods. They also improve on cis-only approaches by removing spurious regulatory relationships that are not supported by TF expression, thereby increasing model precision. Among the first cis- and trans-GRN inference methods are SCENIC+ \cite{bravo_gonzalez-blas_scenic_2023} and LINGER \cite{yuan_inferring_2024}. SCENIC+ links TFs and enhancer regions to a gene by using random forest regression, assuming non-linear relationships between regulators (TFs and enhancers) and target gene expression. Regulatory relationships are inferred using importance analysis. LINGER uses a similar per-gene regression model, where non-linearity is modeled as a two-layer fully connected neural network between TFs and each target. LINGER additionally assumes that population, cell-type and individual cell-specific GRNs are not independent and have shared structure and  that can be inferred by fine-tuning a starting GRN estimated from bulk data. Each of these networks is composed of both cis- and trans-regulatory elements which are inferred with SHAP scores. More recently, scTFBridge \cite{wang_sctfbridge_2025} was developed, which utilizes a multi-modal variational auto-encoder framework to produce shared and modality-specific (e.g., scRNA-seq, scATAC-seq) latent representations. The ATAC-seq decoder is constrained by known TF-motif associations, inferring a latent activity profile of a TF. Cis- and trans-GRNs are inferred using regression of latent activities to downstream peak activity or target expression. scTFBridge was shown to have improved performance compared to SCENIC and other methods indicating the utility of their disentanglement framework. However, scTFBridge relies on known TF-motif relationships, whereas LINGER can incorporate additional TFs. A common theme across all methods was to use a predictive model of  gene expression (or activity) prediction and use model interpretability tools (e.g., feature importance in Random Forests, SHAP values) to obtain regulatory relationships between regulators and target genes. 

\section*{From correlation to causation: Modeling perturbations with GRNs}

\marginnote{
\textbf{Perturb-seq}\\
a scRNA-seq approach assessing genetic perturbations via pooled screening of gene expression profiles and guide RNA identities in individual cells
}[-1.5cm]

\begin{figure}[htbp]
\includegraphics[width=\textwidth]{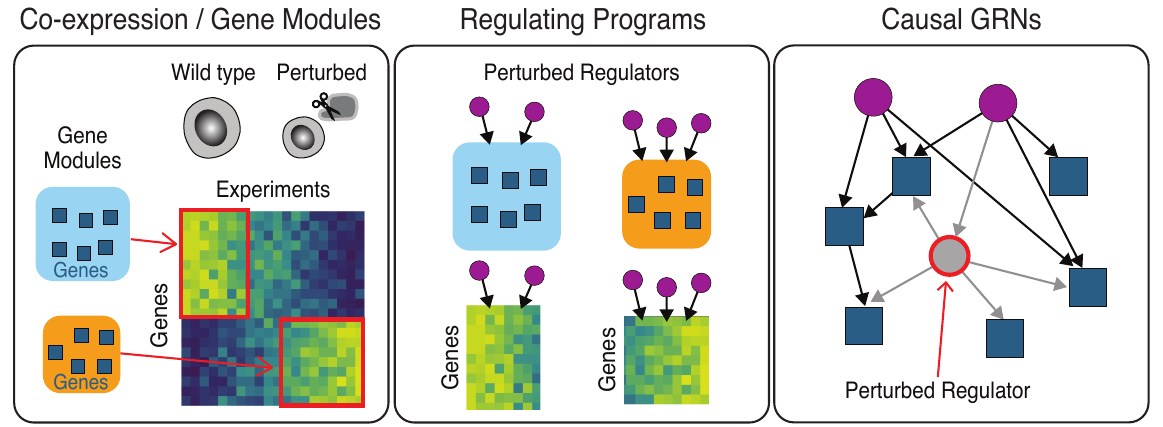}
\caption{\textbf{Inference of causal relationships in GRNs from high-throughput perturbation data, such as Perturb-seq.}  Approaches can be grouped into defining gene expression programs (Co-expression/modules), inferring regulators for the programs (Regulating Programs), inferring causal TF-target interactions (Causal GRNs). Circular nodes: transcriptional regulators; square nodes: target genes.}
\label{fig4}
\end{figure}

With the availability of large-scale perturbation techniques such as Perturb-seq \cite{dixit_perturb-seq_2016} and CROP-seq \cite{datlinger_pooled_2017}, an important emerging direction is the inference of causal interactions. These assays perturb numerous loci at a time and profile high-dimensional molecular readouts like genome-wide gene expression levels. In this section, we review key methodologies for inferring causal GRNs from such perturbational 
data. There are several reviews that cover the different types of experimental techniques of perturbation \cite{bock_high-content_2022} and general classes of computational problems that arise while analyzing these types of data \cite{gavriilidis_mini-review_2024, rood_toward_2024, sanguinetti_network_2019}. We focus on methods for three main tasks (\textbf{Figure \ref{fig4}}): (i) assessing  the effect of perturbing a single regulator, (ii) detection of gene programs, and (iii) detection of causal gene regulatory relationships. As a pre-processing step, quality control in perturb-seq data analysis typically involves filtering cells for sufficient depth,  guide detection, and assignment of guides to cells \cite{adamson_multiplexed_2016, replogle_combinatorial_2020}. 

\subsection*{Assessing of the effect of perturbing a regulator}
To assess the impact of perturbing a single gene, most analysis pipelines follow a ``differential expression'' (DE) type of analysis, comparing the expression profile in a targeted population of cells  to a control cells. These methods require the identification of cells with sufficient perturbation of the targeted gene. Several tools offer the identification of perturbed cells as part of preprocessing steps such as MUSIC \cite{duan_model-based_2019}, PerturbDecode \cite{geiger-schuller_systematically_2023}, and Mixscape \cite{papalexi_characterizing_2021}, or as part of a regression framework allowing for various technical and biological sources of variation in a gene's expression level (e.g., MIMOSCA \cite{dixit_perturb-seq_2016}). MUSIC assumes that for a targeted cell to have sufficient perturbation, its median similarity to all other targeted cells must not be less than the control cells. A perturbation is filtered if most of its targeted cells are more similar to controls. 

PerturbDecode's preprocessing assumes that the impact of perturbation from different guides for the same targeted gene is similar based on a pairwise Pearson correlation cutoff of the effect size on each measured gene. It filters all guides if they do not induce similar responses based on the minimal Person correlation. It then identifies ``impactful guides'' and ``impacted cells'' based on regression coefficients in the guide-by-gene effect size matrix built from fitting a negative binomial linear mixed-effects model that regresses the expression of each gene on the detected guides. The regression model also captures  technical variables as fixed-effect covariates and cell states defined by Leiden clustering as random effects, which is based on the assumption that the perturbation does not affect the cell state definition. 

Mixscape normalizes gene expression using control cells  in the same knn neighborhood as a targeted cell. It thus assumes that control cells may  exhibit different background distributions and normalization of the targeted cell should respect the knn neighborhood. For each guide, it then computes a similarity metric for each guide-positive cell by comparing the cell's normalized gene expression profile to the mean change in these cells' expression profile relative to control cells. A Gaussian mixture model with two mixtures, one for the foreground (perturbed) and one for the background population (unperturbed), is fit to the resulting metrics for filtering cells that ``escape'' perturbation. In contrast to Mixscape, which assigns cells to binary states, recent methods such as Mixscale \cite{jiang_systematic_2025} and perturbation-response score \cite{song_decoding_2025} use continuous scores to capture heterogeneous cellular responses to perturbation in CRISPRi datasets. These scores can be used to weight cells in a multivariate regression to identify DE genes. 

\subsection*{Detecting gene expression programs} 
Gene expression programs are defined as co-expressed gene modules and their  regulators (targeted genes), and capture the perturbation impact of multiple regulators as opposed to individual regulators.
The most common approach for identifying gene expression programs is based on factor analysis, which assumes that gene expression programs are captured by low-dimensional latent factors, each factor corresponding to a common effect across groups of genes forming the gene program. In this framework, the perturbation profile of a cell   is assumed to be a linear combination of low-dimensional gene expression programs. Furthermore,  a gene is assumed to participate in a small subset of these factors. 
Several methods have been developed to estimate gene expression programs, including D-SPIN \cite{jiang_d-spin_2023}, MUSIC \cite{duan_model-based_2019}, PerturbDecode \cite{geiger-schuller_systematically_2023}, and GSFA \cite{zhou_new_2023}.

D-SPIN assumes that there is no overlap between gene sets associated with each program, and uses orthogonal NMF to identify gene programs. D-SPIN uses a Markov random field (MRF) where nodes are gene programs (or genes for small datasets), with their values discretized into one of three states -1 (silenced), 0, and 1 (expressed). The structure of the MRF is inferred by optimizing the overall likelihood or pseudo likelihood of the data and provides the dependency structure among the programs (or genes). D-SPIN's MRF model additionally estimates a ``perturbation response vector'' for each perturbation such that selected programs in the perturbed samples are more likely to be in a certain state, which can be interpreted as the effects of a perturbation on nodes. To identify important regulatory interactions that mediate the effect of a perturbation on a specific gene or gene program, D-SPIN reports an ``edge sensitivity score'' for each individual edge in the GRN by comparing the change before and after removing the edge in the difference of marginal probability of the gene or program being expressed (state 1) and silenced (state -1).

GSFA, like D-SPIN uses matrix factorization to define the gene expression programs, but more explicitly models the impact of perturbations on the latent factors. Specifically, it assumes that the expression matrix factorizes into the product of the design matrix  connecting a guide RNA to a cell and the perturbation-by-gene matrix which itself is a product of the effect size matrix ($\beta$) and the loading matrix ($W$), as ($\beta W^T$). The effect is summarized by sampling from the posterior of the entry in the perturbation-by-gene matrix corresponding to the perturbation-gene pair.

MUSIC assumes expression is generated from a probabilistic topic model, each topic representing an expression program and uses  latent Dirichlet allocation (LDA) to get a topic distribution for each perturbation. Topics are interpreted based on enrichment analysis of Gene Ontology (GO) terms of DE genes, but do not explicitly link perturbed gene to a topic. MUSIC provides ways of prioritizing perturbations by comparing the topic distribution of each perturbation against the control. 

\marginnote{
\textbf{Topic modeling}\\
a statistical method in natural language processing (NLP) that identifies core topics (distributions of words) within document sets, modeled as mixtures of topics, by analyzing word co-occurrence \\[0.2cm]
\textbf{Latent Dirichlet allocation (LDA)}\\
a generative probabilistic model that classifies documents into topics by assigning word probabilities, with each document represented as a mixture of topics
}

Unlike the above methods, PerturbDecode uses clustering to define groups of ``impactful guides'' into modules by the Leiden algorithm on a k-nearest neighbors (kNN) graph constructed from the top 50 principle components of the guide-by-gene effect size matrix. To group genes by their co-regulation by guides, it performs the same previous step but uses the transpose of the effect size matrix. The transpose of the effect size matrix is also used for independent component analysis (ICA) for the identification of pathways and the associated genes and guides. 

\subsection*{Inferring causal gene regulatory relationships}
The third, and perhaps most challenging, analysis is the causal discovery of regulator target gene interactions. Methods that attempt to perform causal discovery from observational data have been extensively studied and applied to GRN inference (see reviews by Glymour \textit{et al.}\cite{glymour_review_2019} and Vowels \textit{et al.}\cite{vowels_dya_2023}). Here we select recent causal discovery methods for learning GRNs incorporating CRISPR-based interventional data. These methods can be grouped into those that directly infer TF-gene interactions and those that detect TFs associated with gene programs, represented as latent factors. 

Dotears \cite{xue_dotears_2023} is an example of a method from the first category, which  assumes that the causal model is represented by a linear structural equation model (SEM) and all nodes in the network are perturbed, but atomically (only one gene is perturbed at a time). Dotears can explicitly use the interventional data as ``hard'' interventions to estimate an improved variance structure that can produce more accurate causal interactions, as shown in simulations. A recent benchmarking study, ``Causal bench'' \cite{chevalley_causalbench_2023}, however showed that methods explicitly incorporating perturbations were not necessarily better than methods which relied only on observational data.

The second class of methods that aim to detect causal interactions at the gene program level, fall under the general category of ``causal representation learning'', where causal regulators may even be latent factors. Several approaches have been developed for this class of methods including Discrepancy-based VAE \cite{zhang_identifiability_2023} and DCD-FG \cite{lopez_large-scale_2022} described before.  Discrepancy-based VAE  encodes observational, unperturbed expression data into latent exogenous terms via a variational autoencoder (VAE) and constructs latent variables representing gene programs through a deep structural causal model, which are used to reconstruct the original data. In addition, it encodes perturbations that act on the latent gene programs and their causal interactions, which results in decoding a counterfactual sample that can be compared with the real interventional data for model training. Perturbations are then grouped by which latent gene program they affect. Another approach, DCD-FG \cite{lopez_large-scale_2022} assumes a low-rank structure of causal interactions among genes grouped as programs but uses a factor graph model instead of MRF. In addition to having a low time complexity, DCD-FG's factor graph representation outputs factors that connect genes in the learned graph and are conveniently interpreted as programs. 

\section*{Interpreting and utilizing GRNs for knowledge discovery}
A central goal of tools for GRN interpretation is the identification of key regulators, functional modules, or subnetworks, to gain biological insights into the regulation of a particular biological process and potentially inform discovery of therapeutic targets in biomedical research. Typically, the most direct strategy involves manual inspection of network components informed by prior biological knowledge. However, this approach becomes impractical when analyzing genome-scale GRNs, or when comparing multiple networks across diverse biological conditions. Computational extraction of network modules  from a single inferred GRN using graph clustering algorithms can assist with downstream interpretation.  Such methods include the PageRank algorithm that used from Epoch \cite{su_reconstruction_2022} or the Granger causality test from NetID \cite{wang_scalable_2024}. 
Interpreting multiple networks is often complicated by the large number of predicted interactions, which vary across cell types and states. To address this issue, several studies have proposed the analysis of differential edge patterns across GRNs constructed under distinct biological contexts. Notable examples include scMTNI \cite{zhang_inference_2023} and LocaTE \cite{zhang_learning_2023}, which identify recurrent regulatory modules by clustering edge adjacency matrices to reveal patterns of edge presence that may reflect underlying biological heterogeneity.

Beyond network structure comparisons, there are also quantitative methods  which provide statistical metrics to prioritize the key components of GRNs. For example, Differentiation Genes’ Interaction Enrichment (DGIE) score, introduced in the scPADGRN \cite{zheng_scpadgrn_2020} study, is a metric designed to quantify differences in inferred dynamic network states over time by measuring the recapitulation rate of known pathway genes. This is essentially a fold enrichment score for a given known pathway, defined as a ratio of counts of binary edges interconnecting genes within a specific pathway over all genes in a state-specific GRN. This score was calculated for identified dynamic GRNs in a biological pathway of interest, such as cell differentiation, to evaluate relevance to the pathway while comparing state-specific GRNs. 

DREAMIT \cite{maulding_associating_2024} is another computational method developed to quantitatively evaluate inferred regulatory interactions within GRNs by incorporating trajectory information. It uses pseudotime data to model gene expression levels for each gene by defining discretized bins along the trajectory. The average values per bin are then regressed to a spline for integrating information from neighboring bins to further smooth expression changes in pseudotime. DREAMIT quantifies the transcription factor (TF)-target relationship by computing similarity measures between them and tests the significance of these relationships through a statistical comparison between the distributions of similarity values from given target genes for a TF and from random targets. Despite its limitations, such as strong assumptions about the correlation between TFs and their targets, DREAMIT offers an intuitive approach for prioritizing regulatory interactions and testing the significance of networks along a cell trajectory.

MIRA \cite{lynch_mira_2022} is another approach that incorporates trajectory information in assessing GRNs. Originally designed to analyze the spatiotemporal dynamics of single-cell transcription and chromatin accessibility, MIRA leverages cell-level topic modeling based on variational autoencoder neural networks and gene-level regulatory potential (RP) modeling. By refining cell state trees based on pseudotime, MIRA identifies key regulators and gene modules governing trajectory branch points. Notably, MIRA quantifies regulatory influence by distinguishing between local or nonlocal chromatin accessibility on gene expression. This is achieved through the computation of local (near the transcriptional starting site, or TSS) and nonlocal (far from TSS) chromatin accessibility-influenced transcriptional expressions, termed ``LITE and NITE,'' based on the expanded gene RP modeling. This method evaluates the effects of chromatin accessibility (scATAC-seq) effects on gene expression, enabling the identification of discrete gene modules governed by cis- or trans-regulatory interaction. The comparison of these two metrics in each cell across cell state trajectories, termed ``chromatin differential,'' reveals how distinct regulatory units govern fate commitment and terminal identity. This type of comparison is particularly beneficial for detailed analysis and evaluation of GRNs derived from the integration of gene expression and chromatin accessibility.

Finally, methods like CellOracle \cite{kamimoto_dissecting_2023}, introduced for cell-type‑specific trans-GRNs can be used to prioritize important regulators through {\em in silico perturbations}. In the original study, CellOracle inferred 24 GRNs from a 2,730‑cell mouse myeloid hematopoiesis atlas and accurately reproduced both systematic and stage‑specific knockout phenotypes using perturbation scores and Markov random‑walk simulations. A large‑scale application to zebrafish embryogenesis, involving {\em in silico} perturbation of 232 TFs across a 38,731‑cell atlas, similarly recapitulated known developmental defects and uncovered new regulators validated experimentally. Across these systems, perturbation scores consistently prioritized key TFs, with most top‑ranked candidates corresponding to established regulators.

\section*{Use cases of GRN inference}
We now highlight several case studies of GRN inference on single cell multi-omics datasets for developmental and disease contexts using SCENIC+ \cite{bravo_gonzalez-blas_scenic_2023}, one of the first frameworks to integrate both cis- and trans-regulatory information. SCENIC+ was applied to a developing human neocortex \cite{wang_molecular_2025}, single cell Multiome dataset, which profiled  232,328 nuclei across 33 cell types to infer enhancer driven regulons (eRegulons). Consensus chromatin peaks were called per cell type to generate a peak by nucleus matrix, followed by topic modeling to identify regulatory topics. Candidate enhancer regions were defined through topic-based selection and differential accessibility (DA) testing and then linked to TFs via motif enrichment. SCENIC+ subsequently identified TF-region-gene triplets, yielding 582 high confidence eRegulons. Cell type and age-specific eRegulon activities were quantified using accessibility  and expression based AUC scores, confirming expected positive (activators) or negative (repressors) correlations with their targets. This activity profiling highlighted distinct GRNs in cortical progenitors and differentiated neuronal populations. Trajectory inference with Slingshot \cite{street_slingshot_2018} further examined lineage-specific eRegulon dynamics by identifying bifurcation associated regulators that drive neuron subtype diversification through trajectory based differential expression (DE) analysis. 

Complementary work in a mouse visual cortex Multiome atlas spanning embryonic day 11.5 to adulthood \cite{gao_continuous_2025} applied a modified SCENIC+ workflow. From 882,075 chromatin accessibility peaks, they first identified DA peaks across all cell‑type subclass by age groups through pairwise statistical testing. These DA peaks were then clustered into peak modules based on shared cell type specificity and temporal patterns. Potential TF regulators for each module were inferred using differential motif enrichment across modules, and candidate peak-gene links were defined by high accessibility-expression correlations across cell type subclass and age profiles. Using the SCENIC+ framework, TF-peak-target triplets were assembled and ranked with confidence scores. Finally, activating or repressing relationships were inferred using TF-peak and gene correlations and the sequential timing of TF, peak, and gene activation. Applying this framework revealed shared and subclass-specific early and late GRNs across major neuronal and glial lineages.

Beyond developmental systems, SCENIC+ has been applied to disease-related contexts to characterize dynamic GRNs in pathological conditions. One such study used SCENIC+ on a single‑nucleus Multiome data profiling the injured mouse spinal cord \cite{zamboni_regulatory_2025}. The authors first applied ArchR \cite{granja_archr_2021} for candidate peaks calling, followed by ChromBPNet \cite{pampari_chrombpnet_2025} on DA regions to identify thousands of cell‑type‑specific injury‑responsive enhancers (IRENs) and learned their sequence level motif syntax. SCENIC+ was then applied to reconstruct lineage specific eRegulons by linking TFs to IRENs and their target genes. This revealed coordinated interactions between AP-1 family factors and lineage-specific TFs in selecting distinct IRENs in each glial subtype. 

In another study, SCENIC+ was applied to a single‑cell Multiome dataset of mouse tumor model of neuroblastoma \cite{xu_single-cell_2025},  to reconstruct cell type‑specific GRNs. Peaks were first called from tumor state-specific pseudo bulk accessibility profiles and merged into a consensus peak set. Topic modeling then identified co-accessible region sets, which were then used for downstream DA analysis. SCENIC+ eRegulon inference and the AUCell tool, which ranks cells by activity of eRegulons, identified 353 active TFs. Comparison with 422 state associated TFs from 49 patient tumors yielded 121 core regulators, which further grouped into key tumor cell-state programs. 

Together, these studies highlight that combining chromatin accessibility, and gene expression, within a cis-trans-GRN inference framework can provide critical mechanistic insights into  dynamics of developmental  and disease-specific processes.

\section*{Future outlook}
In this review, we covered current problems and computational approaches to infer gene regulatory networks from single cell multi-omic datasets. Going forward we envision methodological advances to address three major directions.

\subsection*{Incorporation of chromatin state and three-dimensional genome organization in dynamic GRNs}
One area of methodological development is more explicit use of chromatin state, including accessibility, in the inference of GRN dynamics. Majority of approaches for modeling GRN dynamics have done so at the trans level by incorporating accessibility to construct a prior for the network inference task \cite{burdziak_sckinetics_2023, wang_dictys_2023, zhang_inference_2023} and capture only cell cluster-level dynamics. The DynaVelo method goes a step further by leveraging neural ODEs to capture continuous dynamics of the cell state in the latent space, however, the GRNs are inferred as a model interpretation step. Furthermore, the accessibility profile of a cell is a summarization of binding sites of each TF (chromVar deviation scores)  which loses locus-specific information that could be important for the cell state dynamics. Extending these approaches to more explicitly model the accessibility-informed GRN structure could provide a more comprehensive and precise view of the GRN dynamics and cell state. 

\subsection*{Integration of promoter proximal and long-range regulatory interactions in for cis- and trans-GRN inference}
Another direction where GRN inference methods could be extended is to incorporate  3D genome architecture to infer  dynamic  cis- and trans-GRNs. In particular, the availability of multi-ome datasets has enabled better modeling of cis-GRNs relating enhancers to genes. However, these models have focused largely on promoter-centric signal and do not explicitly model TF expression  to predict the expression of the target gene as is done in trans-GRNs. With the increasing availability of single cell (scHiC) data together with gene activity level measurements (e.g., mRNA) in the same cell, new methods could leverage these data to capture cis-trans dynamic GRNs. A major challenge that will need to be overcome, is integration of scHic, scRNA-seq and accessibility to define cell state, pseudotime or velocity, while being allowing for the complementarity of these assays to define cell states and types. 

\subsection*{Benchmarking of causal GRNs with multi-modal measurements post perturbation}
Evaluation of GRNs remains challenging, largely due to limited gold standard networks. Much of the gold standards that exist are built from bulk assays including a combination of TF knock-out and TF-binding data (ChIP-seq and CUT\&Tag). The availability of Perturb-seq assays has significantly scaled up downstream validation. However,  Perturb-seq measures transcription readouts and there is only limited measurements of accessibility or other readouts at the single cell. Developing such multi-modal readouts post perturbation while keeping the assays scalable and cost-effective will require advances in experimental and computational methods. 

Accordingly, metrics to assess the validity of GRNs would need to be improved to consider more general properties of the structure and function of GRNs. Current structural metrics focus on individual edge and node level agreement with ground truth. However, this poses the question of comparing networks at different resolutions, capturing higher-order relationships at the level of subnetworks and modules. Graph neural network-based representations of GRNs could be an interesting direction of work to enable these general comparisons. At the function level, GRN inference models would need to predict the global high-throughput readouts post perturbation,  requiring faithful modeling, e.g., via generative models,  that could be sampled under ``\textit{in silico} perturbations''. 

\subsection*{Inherently interpretable causal GRNs with high predictive power}
The availability of large-scale perturbation data has enabled the development of shallow and deep learning methods that can accurately predict the impact of perturbations \cite{ahlmann-eltze_deep-learning-based_2025,lotfollahi_scgen_2019,roohani_predicting_2024}. However, the mechanisms by which these perturbations affect the structure of gene regulatory networks (GRNs) are often inferred through model interpretation. Learning GRNs from both perturbation and observational data in a scalable and cost-effective manner remains a significant challenge. Early work in causal representation learning, such as Discrepancy VAE \cite{zhang_identifiability_2023} and DCD-FG \cite{lopez_large-scale_2022}, offers promising avenues for further advancement. Extending these approaches to incorporate additional modalities and dynamic behaviors, along with systematic benchmarking of methods, is an important direction for future research.

\section*{DISCLOSURE STATEMENT}
The authors are not aware of any affiliations, memberships, funding, or financial holdings that might be perceived as affecting the objectivity of this review. 

\section*{ACKNOWLEDGMENTS}
This research was supported in part by the National Institutes of Health (NIH) NHGRI grant 5R01HG012349 (JS, SH, SR) and NIGMS grant 1R01GM144708 (YL, SR). SHS is supported by the NIH training grant (5T15LM007359) and the H.I. Romnes Faculty Fellowship awarded to SR. 
\clearpage

\nolinenumbers
\bibliography{references}
\bibliographystyle{abbrv}

\end{document}